\documentclass[doublecol]{epl2} 

\usepackage{amssymb,amsmath}
\usepackage[ansinew]{inputenc}
\usepackage{enumitem}   
\usepackage{amssymb}
\usepackage{graphicx}

\title{Aerodynamics of the square-back Ahmed body under rainfall conditions}
\shorttitle{Aerodynamics of the square-back Ahmed body under rainfall conditions} 

\author{N. Mazellier\inst{1} \and M. Obligado\inst{2}}
\shortauthor{N. Mazellier \etal}

\institute{                    
  \inst{1} University of Orl{\'e}ans, INSA-CVL, PRISME, EA 4229, 45072, Orl{\'e}ans, France\\
  \inst{2} Universit{\'e} Grenoble Alpes, CNRS, Grenoble-INP, LEGI, F-38000, Grenoble, France
}
\pacs{47.55.Kf}{Multiphase and particle-laden flows}
\pacs{47.27.Vf}{Wakes}
\pacs{47.27.Ak}{Turbulent flows, fundamentals}

\abstract{
We report an experimental investigation about the aerodynamics of a simplified road vehicle, the so-called square-back Ahmed body, under rainfall conditions. A particular emphasis is put on the evolution of the body base pressure distribution with respect to the operating conditions. It is found that rainfall significantly damps both mean base pressure drag and wake dynamics in comparison to dry conditions.}

\begin{document}

\maketitle
\section{Introduction}
Road transport dominates the freight and passenger mobility sector. However, this strategic sector faces considerable challenges to conjugate the increasing need for freedom of movement with sustainable growth. To promote the development of low environmental impact transport, national and international laws set mandatory emission reduction targets for the automotive industry. 

In this context, the so-called Ahmed body was proposed as a generic bluff-body representative of road vehicles \cite{ahmed1984}. Over the past decades, large efforts have been invested by the scientific community to investigate and control the complex aerodynamics of simplified road vehicles at the laboratory scale \cite{grandemange2013,barros2017,mcnally2019}. However, the role of external conditions remains largely unexplored. Recently, the influence of background turbulence on the aerodynamics of square-back Ahmed body was investigated \cite{burton2021,passaggia2021}. These studies emphasized that both drag coefficient and wake dynamics at low frequency are strongly sensitive to the turbulent properties of the incoming flow. Most of these studies were performed in single-phase conditions. 

Very few studies were devoted to the aerodynamics of road vehicles under heavy rain \cite{gaylard2014,patel2022,raeesi2023}. More generally, very little attention has been paid to the coupling of inertial particles with aerodynamics, and only a handful of studies are currently available in the literature  \cite{cao2014effects,wan2004aerodynamic}. This is a remarkable omission, as turbulent wakes charged with inertial particles (such as rainfall) have also important environmental and industrial applications. For instance, recent works have found that the presence of inertial particles can modify the structure and persistence of the turbulent wake of scaled wind turbines \cite{smith2021dynamic,travis2022} and to  affect the performance and aerodynamic stall of aircrafts \cite{sheidani2022study}.

In this work, we report recent experimental investigations of the square-back Ahmed body carried out under rainfall conditions. The main objective is to determine the influence of climatic conditions on the aerodynamical characteristics of this simplified road vehicle. For such aim, water droplets with different volume fractions are injected in a wind tunnel and the pressure distribution in the back wall of a square-back Ahmed body is studied. Our results show that rainfall has a significant influence in the bistable nature of the turbulent wake generated by the body and that it also reduces the pressure drag on it when compared to dry conditions.

\section{Experimental set-up}

\begin{figure*}
\centering
\includegraphics[width=0.9\textwidth]{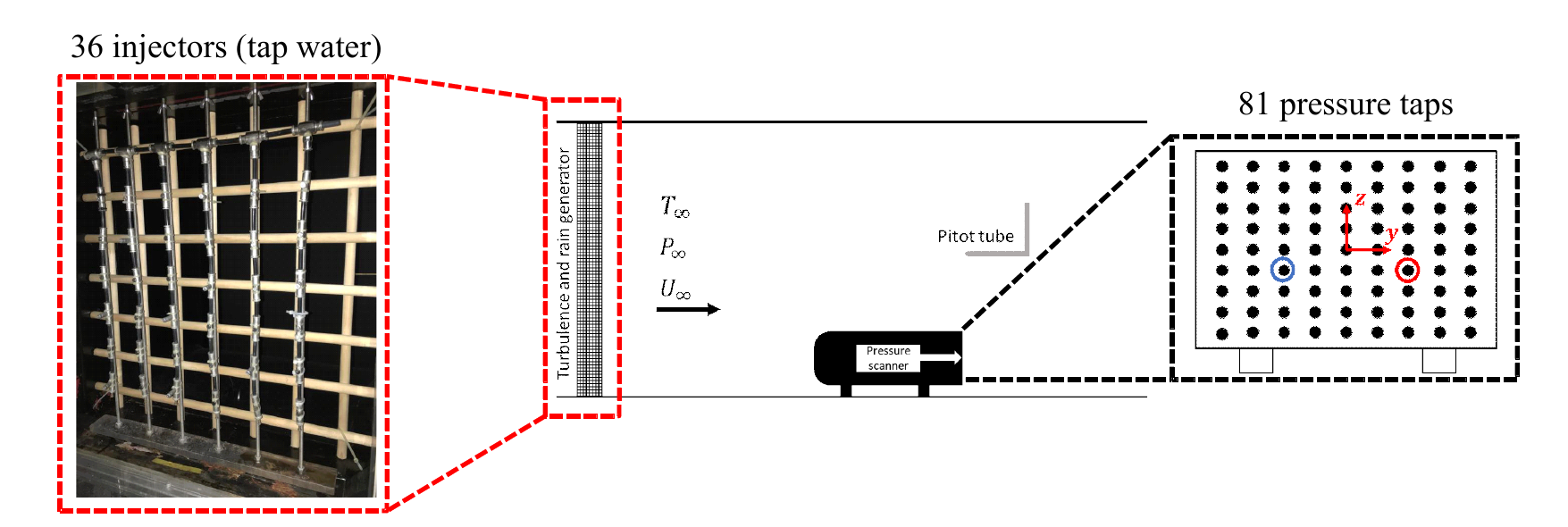}
\caption{Sketch of the experimental setup, showing the passive grid and the injection system (left picture). The distribution of the pressure taps is displayed in the right insert together with the coordinate system. The pressure taps indicated by the blue and red circles are those used in figure \ref{fig4} to illustrate the switching of the wake from left to right and vice versa.}
\label{fig1}
\end{figure*}

Experiments were carried out in the Lespinard wind tunnel at LEGI laboratory (Grenoble, France). The test section of this closed-loop wind tunnel is a $0.75\times0.75 \mbox{m}^2$ square extending over 4 m (see figure \ref{fig1}).

\begin{figure}
\centering
\includegraphics[width=0.4\textwidth]{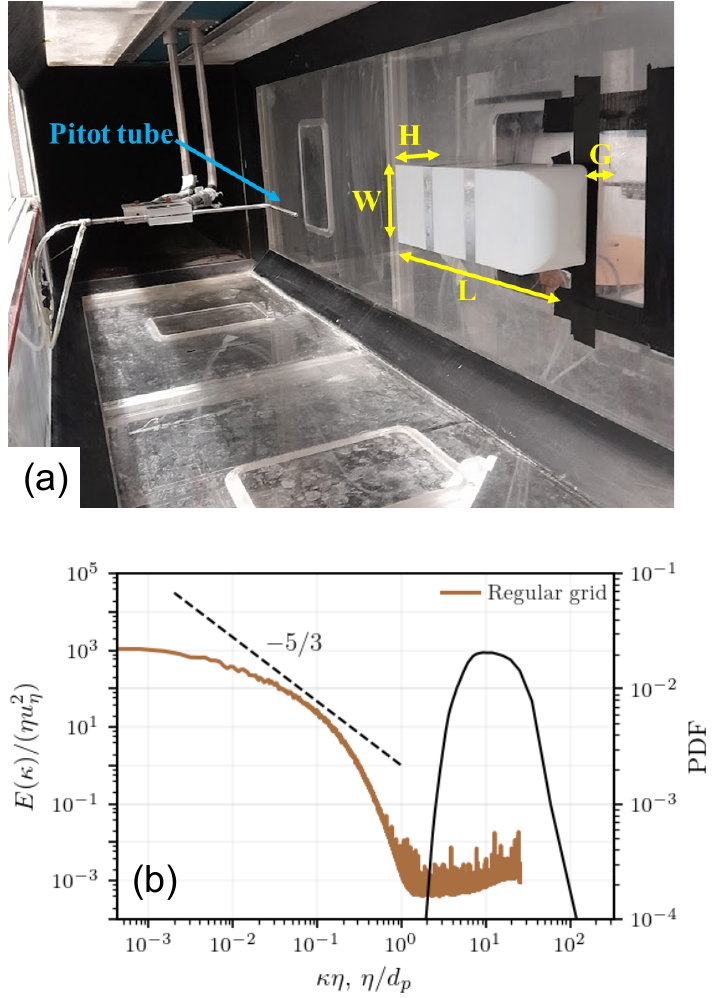}
\caption{(a) Picture of the Ahmed body positioning inside the wind tunnel. The pitot tube can be observed downstream of it. (b) Velocity spectra obtained using hot-wire anemometry obtained $\sim27M$ away from the passive and injection grids (brown line, the hot-wire anemometry system is described in \cite{ferran2022experimental}). The black solid line corresponds to the probability density function (PDF) of droplet size obtained using a Spraytec (see \cite{sumbekova2017preferential} for more information about how data was collected).}  
\label{fig2}
\end{figure}

A 3D-printed square-back Ahmed body, with dimensions [$L$, $W$, $H$] = [381, 142, 105] mm, was set within the test section 2.5 m downstream of its inlet (figure \ref{fig2}a). The ground clearance of the model $G$ was set to 18 mm. It is worth noticing that, the model was installed on a vertical wall of the wind tunnel. This positioning made it possible to limit any problems of blockage due to water stagnation underneath the model during the tests.

As illustrated in figure \ref{fig1}, water droplets were injected with a regular grid composed by 36 nozzles (0.4 mm diameter) at the test section's inlet. A 150 bar pump supplied water to the spray grid, producing a uniform spray of poly-disperse water droplets with most probable diameters between 40-50 $\mu$m (figure \ref{fig2}b). Three different volume fractions where tested (defined as the ratio between the volume of liquid and air at the injection), $\phi_v = 2.5\times10^{-6}, 3.5\times10^{-6} \mbox{ and } 4.6\times10^{-6}$. More information about the injection system can be found in \cite{sumbekova2017preferential,ferran2022experimental}.

A passive grid (mesh size $M \approx H$) was located 15 cm upstream of the spray grid to mix any effects and inhomogeneities from the injection system. Measurements in the back wall of the Ahmed body were performed at approximately $27 M$ downstream the static grid, where turbulence has been found to be nearly homogeneous and isotropic and the combined grids produce a turbulence intensity in the order of 2.5\%. In all cases, the particle size remained below the Kolmogorov length scale of the turbulent flow (figure \ref{fig2}b, see \cite{ferran2022experimental} for further details).

 The free-stream velocity $U_\infty$ was recorded by means of a Pitot tube located downstream of the model. The results reported here have been obtained for $U_\infty \approx 8.3$ m/s, leading to a Reynolds number $Re_H$ ($\sim U_\infty H/\nu$ with $\nu$ the kinematic viscosity of air) around $5.8 \times 10^4$. In the following, $U_\infty$ and $H$ are used for normalisation purposes.
 
\section{Methodology}

Since the near wake of the square-back Ahmed is characterised by a massively separated region, its aerodynamics is intimately linked to the base pressure evolution. Accordingly, in this study the influence of rain on aerodynamics is investigated by monitoring the pressure distribution at the base of the model. To this end, 81 pressure taps were installed at the body base (see figure \ref{fig1}). They were equally distributed over the body base  in a rectangular mesh ($\Delta y = 16$ mm and $\Delta z = 11.5$ mm) and probed using 3 microDAQ pressure scanners with an acquisition rate of 100 Hz. The static pressure $P_\infty$ collected by the Pitot tube was used as reference to define the pressure coefficient $C_p = 2\left(P-P_\infty\right)/\rho U_\infty^2$ where $P$ and $\rho$ stand for the local pressure and the density of air, respectively. Besides, during the experiments, the free-stream temperature $T_\infty$ is measured by means of a thermocouple set within the test section.

To appropriately converge statistics, each run lasted around 30 min and was performed in 3 steps. First, for about 6 min, the base pressure distribution was recorded in dry conditions, which are used as a reference. Then, during the second step, the water injectors were activated for about 20 min. Finally, the last step corresponds to the return to dry conditions by stopping the water injection. Before restarting a new try, the wind tunnel is kept running to ensure that the model and the test section are fully dried. 

It is important to note that once the water injection is activated, the wind tunnel is fully saturated, meaning that both free stream pressure ($P_\infty$) and temperature ($T_\infty$) remain constant. However, since humidity can modify both the density of the fluid and the reference pressure in the test section, care should be taken when changing operating conditions from dry to wet. Under dry conditions, the density of air $\rho_a$ is calculated from the ideal gas equation. The density variation under wet conditions is estimated as $\Delta \rho = \phi_v\rho_w/\rho_a$, with $\rho_w$ the density of water. In this study, $\Delta \rho$ remains lower than 0.4\%. For what concerns the measurement of the reference pressure $P_\infty$, a specific protocol has been implemented to prevent uncertainties induced by droplet impact on the Pitot tube. To this end, the Pitot tube is disconnected when water injection is started. Besides, a channel of each pressure scanner connected to the atmospheric pressure $P_{atm}$ in the room is used as reference. Then, data collected under wet conditions are corrected using the offset $\Delta P = P_{atm} - P_\infty$ estimated under dry conditions.

\section{Results}
In the following, we will focus in the main differences observed in the aerodynamics of the Ahmed body in rainfall and dry conditions. Indeed, not only  the near wake is affected in average (as shown in figure \ref{fig3}a), but its temporal dynamics is also strongly impacted by the presence of rain. For the sake of visibility, the time range when injectors were activated (step 2) is highlighted with a green window in figures \ref{fig3}(a) \& \ref{fig4}.

\subsection{Mean aerodynamic performances}

\begin{figure}
\centering
\includegraphics[width=0.45\textwidth]{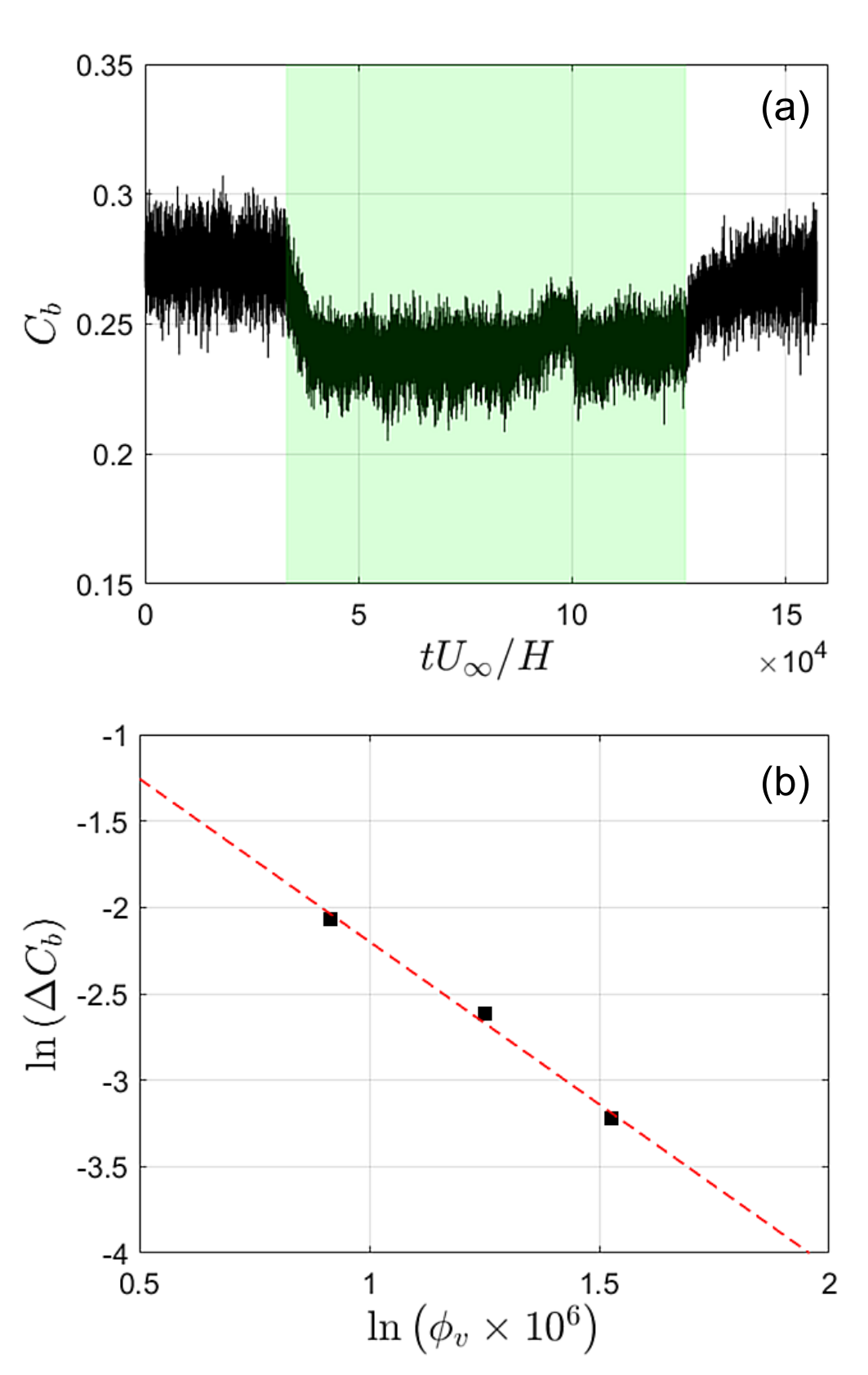}
\caption{(a) Typical time history of the base drag coefficient obtained in dry and rainfall operating conditions. The green rectangle symbolizes the activation of the rain supplier device. While values displayed correspond to $\phi_v = 2.5\times10^{-6}$, the other two volume fractions present similar trends. (b) Variation of $\Delta C_b$ (defined according to equation \ref{eqcb}) with the volume fraction $\phi_v$. The red dashed line represents a power law fitting in the least-mean square sense.}
\label{fig3}
\end{figure}

Figure \ref{fig3}(a) displays a typical time history of the base drag coefficient $C_b$, defined as

\begin{equation}
C_b = -\frac{1}{HW} \iint C_p dS.
\end{equation}

\noindent This physical parameter is used as a metric of drag change when varying the incoming flow properties.  In dry conditions, $C_b$ remains steady around 0.27 in average, a value in good agreement with those reported in previous studies \cite{passaggia2021,zeidan2023near}. As far as two-phase conditions are imposed, one can observed that $C_b$ drops until it reaches a plateau around 0.23, which represents a decrease by 14\% in comparison to the dry conditions. While only the lowest value of $\phi_v$ is shown, all the other ones show a similar behaviour. This might suggest a reduction of drag coefficient but care should be taken since the friction drag, which is not measured in this study, could have been also impacted. Once the rainy conditions are interrupted, $C_b$ returns towards its initial level.

\noindent To further study this effect, we quantify the effect of $\phi_v$ by defining the variation of the base drag coefficient as,

 \begin{equation}\label{eqcb}
 \Delta C_b = 1- \frac{\langle C_b \rangle_{wet}}{\langle C_b \rangle_{dry}},
 \end{equation}

\noindent where $\langle \: \rangle$ denotes time average. Figure \ref{fig3}(b) shows that $\Delta C_b$ decreases with increasing $\phi_v$. At the highest injection rate studied here, $\Delta C_b$ reaches 4\% meaning that base drag reduction is still observed though less pronounced. A previous study \cite{raeesi2023} tested a comparable trend for a DrivAer model. However, the cited work reported drag increase, although the injection rates were comparable to those studied in the present case. This might be related to the properties of the dispersed phase which in their case is characterized by an average droplet size approximately 15 times larger than that the one used in the present work. Accordingly, the Weber number in the experiments carried out in \cite{raeesi2023} is much higher than the one featuring this study implying that the interactions between the incoming rain and the body surface might be very different in nature, making direct comparison irrelevant. This issue is beyond the scope of this study and is therefore left for future works. One can see that, although our study is restricted to 3 operating conditions, the evolution of $\Delta C_b$ is well fitted by a power law such that $\Delta C_b \sim \phi^{-\alpha}$ with $\alpha=1.88$. This trend should be confirmed in the future by studying a broader range of injection rate, which requires a significant adaptation of the experimental set-up.

\subsection{Wake dynamics}

\begin{figure}
\centering
\includegraphics[width=0.48\textwidth]{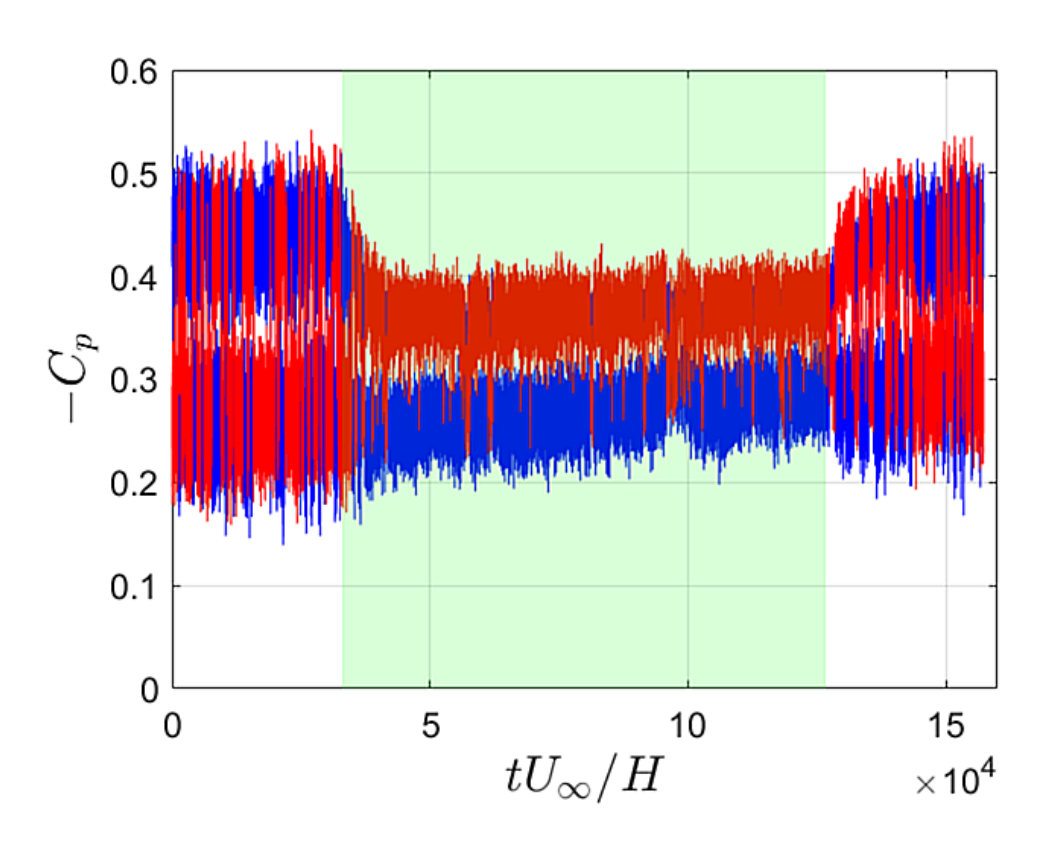}
\caption{(a) Typical time history of the pressure coefficient obtained in dry and rainfall operating conditions. The green rectangle symbolizes the activation of the rain supplier device. While values displayed correspond to $\phi_v = 2.5\times10^{-6}$, the other two volume fractions present similar trends. The pressure coefficient is reported for two sensors located symmetrically around the midspan plane (see figure \ref{fig1} for more details)}.
\label{fig4}
\end{figure}

The wake of the square-back Ahmed body is characterized by a reflection symmetry breaking on a short time scale, which manifests by a random switching of the wake from side to side (in the present work, from left to right (in the direction of $W$), and vice versa since $H<W$). This is evidenced in figure \ref{fig4}, which shows the time history of the pressure coefficient collected on both sides of the base midspan. The location of the pressure taps used here are indicated by blue (left) and red (right) circles in figure \ref{fig1}. Initially, in single-phase flow, it can be easily observed the so-called bi-stable phenomenon induced by the switching of the recirculation bubble from left to right and vice versa. While this behaviour still persists in rainfall conditions, a sharp damping in its amplitude is found. Furthermore, the switching rate is also reduced. Once the rain is interrupted, the initial bi-stable behaviour is quickly recovered. 

\begin{figure}
\centering
\includegraphics[width=0.48\textwidth]{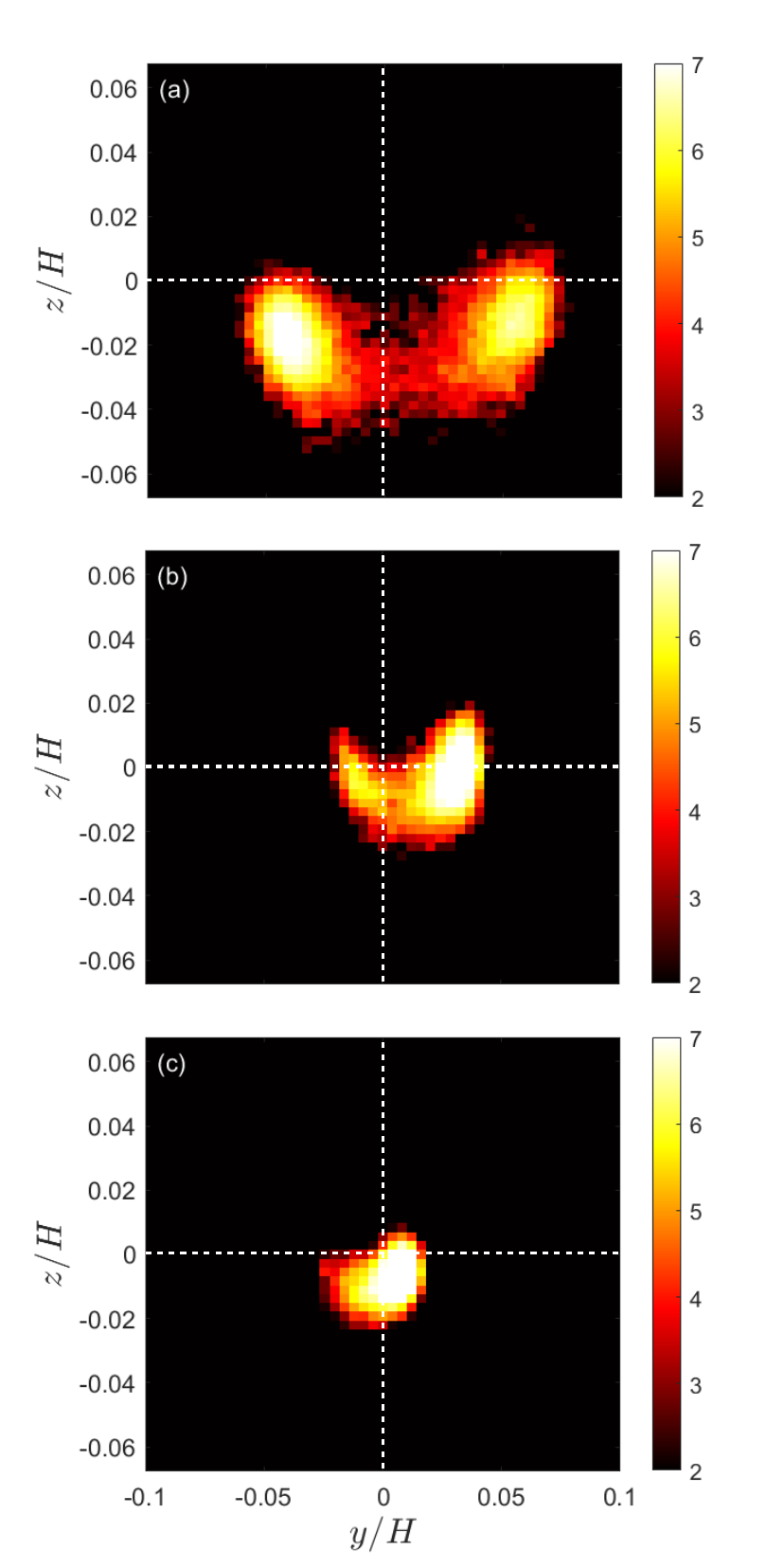}
\caption{Joint-PDF of the base pressure centroid calculated in dry (a) and rainfall conditions: (b) $\phi_v = 2.5\times10^{-6}$ and (c) $\phi_v = 4.6\times10^{-6}$. The intersection of the white dashed lines symbolize the center of the body base.}
\label{fig5}
\end{figure}

The influence of the free-stream conditions is even more emphasised in figure \ref{fig5}, that displays the joint probability density function of the base pressure centroid those coordinates are defined as follows,

\begin{equation}
   \begin{array}{r c l}
      y_c  & = & \frac{1}{C_b} \iint y C_p dS, \\
      z_c  & = & \frac{1}{C_b} \iint z C_p dS.
   \end{array}
\end{equation}


In dry conditions (see figure \ref{fig5}a), the base pressure centroid remains preferentially in two symmetrical locations around the midspan at $y/H \approx \pm 0.05$ and is slightly shifted towards the wind tunnel floor at $z/H \approx -0.02$.

As evidenced in figures \ref{fig5}b and \ref{fig5}c, this switching process is significantly damped for the two-phase flow for which the base pressure centroid departs much less from the base centre. Such damping could be induced by the modification of the boundary layers developing around the body due to the presence of a thin layer of water, which has been observed during the experiments. This lubricating layer may change the wall shear stress as well as the turbulence production, which has been shown to be a key ingredient for the triggering of switching process \cite{hesse2021}. Another (potentially concomitant) explanation is related to the added mass transported within the recirculation bubble in presence of rainy conditions \cite{travis2022,lorite2020}. This may lead to an increase in the energy barrier required for the switching process to be triggered. While such processes cannot be fully addressed in the present experimental set-up, further studies may shed light on their role in the effects observed in our dataset.

\begin{figure}
\centering
\includegraphics[width=0.48\textwidth]{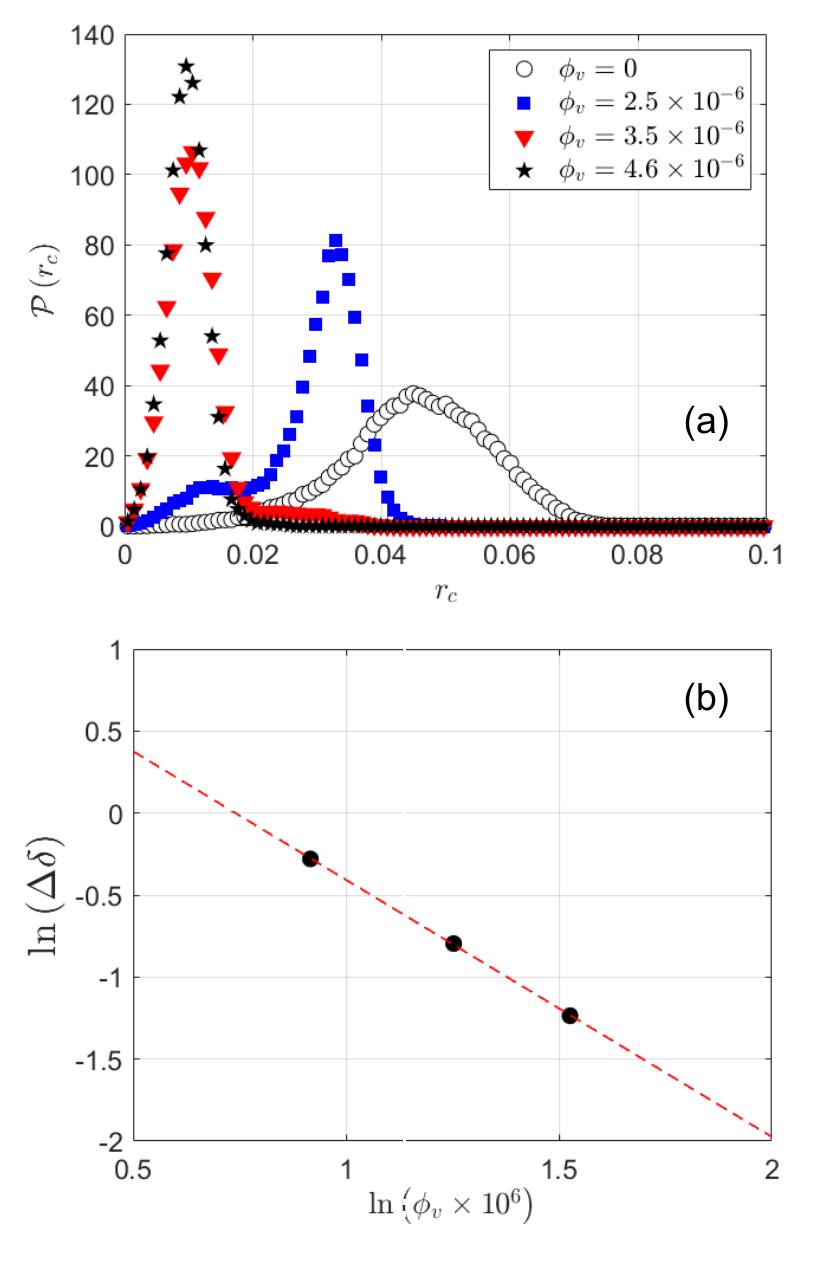}
\caption{(a) PDF of the centroid position, defined as $r_c=\sqrt{y_c^2+z_c^2}$, for all free stream conditions explored. (b) Variation of $\Delta \delta$ (defined in equation \ref{eqr}) with the volume fraction $\phi_v$. The red dashed line represents a power-law fitting in the least-mean square sense.}
\label{fig6}
\end{figure}

A comparison between figures \ref{fig5}b and \ref{fig5}c emphasizes that the dynamics of the base pressure centroid, and accordingly the body wake, is strongly sensitive to the water injection rate. Indeed, the motion of the pressure centroid decreases for increasing $\phi_v$. To investigate further this effect, the PDF of the centroid radius, defined as $r_c = \sqrt{y_c^2 + z_c^2}$, is plotted in figure \ref{fig6}a with respect to the water injection rate. It is found that the effect of the rainfall is twofold: increasing the value of $\phi_v$ not only displaces the value of $r_c$ towards the center, but it also results in narrower distributions, indicating the damping of the bistability. Interestingly, contrary to what is observed for dry conditions, the PDF of $r_c$ obtained for the lowest injection rate (i.e., $\phi_v = 2.5\times10^{-6}$), is characterized by two peaks located at $r_c \approx 0.013$ and $r_c \approx 0.032$, the amplitude of the former being about 8 times lower than that of the latter. For increasing $\phi_v$, the first peak becomes predominant, its position remaining almost unchanged. Besides, the fluctuations $r_c'$ characterizing the excursion of the centroid radius around its mean value decrease with increasing $\phi_v$. To quantify this effect, we introduce the centroid brush $\delta = \sqrt{\langle r_c'^2 \rangle}$, which evolution with respect to the free stream conditions is estimated as follows
 
 \begin{equation}\label{eqr}
 \Delta \delta = \frac{\delta_{wet}}{ \delta_{dry}}.
 \end{equation}
 
\noindent Figure \ref{fig6}(b) shows how $\Delta \delta$ changes with the volume fraction $\phi_v$. Remarkably, a clear trend emerge for this quantity, as $\Delta \delta$ decreases with $\phi_v$. More interestingly, this variation is properly fitted by a power law, with the functional form $\Delta \delta \sim \phi^{-\beta}$, with $\beta=1.57$. We remark that we only tested three values of $\phi_v$ and studies covering a broader range of volume fractions can confirm the trends observed here.

\section{Conclusions}

We report experiments on the influence of rainfall in a square-back Ahmed body under rainfall conditions. Three different values of volume fractions were tested for a Reynolds number, based on the height of the printed body $H$, of $Re_H=5.8 \times 10^4$.

By applying a measuring protocol that alternates dry and rainfall conditions, the influence of the latter could be quantified. First, we found that the base drag is significantly reduced in two-phase conditions. Further experiments measuring the friction drag could complement the present ones and estimate the total drag of the body in such conditions. Furthermore, an analysis of the pressure coefficients and centroids shows that rainfall reduces both the amplitude and frequency of the bi-stability of the wake. 

We also find that, when the void fraction is increased, the centroid position moves close to the center of the body and the base drag coefficient is increased. Further studies exploring a broader range of $Re_H$ and $\phi_v$ would help to confirm the trends observed here, and identify appropriate scaling laws for these quantities.

\acknowledgments
The authors would like to acknowledge St{\'e}phane Loyer (PRISME), Joseph Virone (LEGI) and Vincent Govart (LEGI) for their contributions to wind tunnel measurements.

\bibliographystyle{plain}

\end{document}